# ANALYSIS STUDY OF TIME SYNCHRONIZATION PROTOCOLS IN WIRELESS SENSOR NETWORKS

Salim el khediri[1,2], Nejah Nasri[1], Mounir Samet[1], Anne Wei[2] and Abdennaceur Kachouri[3]

[1]Laboratory of Electrical Engineering and Information Technology, National School of Engineers of Sfax, University of Sfax, Tunisia
`salim.el.khediri@gmail.com`
[2]National Conservatory of Arts and Métiers Paris, France
[3]Higher Institute Of Industrial Systems Gabes, University of Gabes, Tunisia

**ABSTRACT**

*One of the main pervasive problems Wireless Sensor Networks (WSN) encounter is to maintain flawless communication sharing and cooperative processing between sensors via radio links to ensure a reliable treatment of information. Many applications based on these WSNs consider local clocks at each sensor node that need to be synchronized to a common notion of time. In this context, the majority of previous researches were focused on the study of protocols, and algorithms that address these issues in order to resolve synchronization problems. Previous fforts and empirical studies in wireless sensor network (WSN) proposed several solutions (algorithms). The focus of this this paper is to examine and evaluate the most important synchronization algorithms based on the positions of various quantitative and qualitative synchronization protocols for energy-efficient information processing and routing in WSNs.*

## 1. INTRODUCTION

Synchronization in wireless sensor networks is vital aspect of successful and efficient network operations in any business settings, particularly, in military and medical applications, as the latter rely on the data accuracy to make rapid and sound decisions.

Overall, the proposed technique in sensor networks requires that all sensor nodes have a common time scale so that the central unit can coordinate and collaborate between sensors to accomplish their tasks. However, it is difficult to maintain a common time scale for all sensors, so the IEEE 802.15.4 [25] standard has not defined clearly the synchronization mechanisms. The synchronization mechanism is a phenomenon subject to many constraints, which must meet several requirements. These limitations sometimes can be incompatible, such as minimizing energy consumption, reducing the associated costs, and maximizing the quality and accuracy of services provided. The problems of time synchronization have been studied thoroughly in the Internet and local networks. Several technologies such as Global position System (GPS) have been used to provide synchronization in large networks. Other protocols such as Network Time Protocol (NTP) [24] have been developed to keep the clocks ticking on the Internet. However, the time synchronization requirements differ significantly in the context of use of sensor networks. In general, these networks are dense, composed of a large number of sensor nodes. This property makes a lot of difficulties to keep the central synchronization. Energy efficiency is another major problem in synchronization problem due to limited battery capacity of nodes. In this paper, we present the different existing techniques for synchronizing sensor networks.

The various sections of this paper are organized as follows: Section 2 analyzes different network problems. Section 3 presents the various criteria of time synchronization. Section 4 discusses various existing time synchronization algorithms. Section 5 gives our comparison based on the existing synchronization algorithms. Finally, Section 6 contains the conclusion of the paper.

## 2. Network related problem

The intricacies in synchronization problems in wireless sensor network, provides an opportunity for an in-depth analysis research:
First, the clock in different devices must be set at the same reference time. To make this time scale work better, we must synchronize the each clock in node to get a reference time source. So, the local time provided for each node must be the same.
Synchronization plays a crucially important role in wireless sensor networks because it allows the entire system to cooperate and accomplish a complex task of data transfer. For instance, we can cite the coordination the data collected at different nodes, which are grouped into, is a significant result despite the system's node clock-difference, which can begin communicating at different times. In addition, the clock may be modified because of environmental conditions. Second, in order to save energy and increase the lifetime of the network synchronization we can use it appropriately. The sensors can sleep at appropriate times and wake up if necessary. In order to save energy, the nodes must be asleep and awake when time interval ordinates and functions in this respect, we can mention the radio receiver of the node when there are data sent to require precise synchronization between the sensors. Traditional methods of synchronization are not approved for use in the sensor network due to problems of complexity and high power consumption. For example, NTP that works well on Internet to synchronize computers isn't suitable in wireless networks, because it needs a large energy. In addition, GPS can be too expensive to be fixed on low-cost devices since it cannot be available everywhere as well as inside buildings or under water. It should be noted that some middle GPS can't be trusted.

## 3. Criteria of Time Synchronization

In this section will detail the various synchronization problems and network caracterstiques of wireless sensors during operation. The usefulness of the sensor network presence is needed to meet the needs of the users queries by merging the data from each sensor to give a single result. To accomplish this task it becomes necessary for these sensors to agree on a concept such as time. All active sensors (participants) can be wrapped in a common time scale either by synchronizing local clocks in each sensor to the transfer of timestamp (timestamp) has a sensor that arrive with time to its local clock. Assuming various criteria, time synchronization protocols can be identified into different classes

### 3.1. Master-slave versus peer-to-peer synchronisation

- **Master-Slave:** A master-slave protocol assigns one node as master and the other nodes as slaves. These plans reading of the master local clock as a time reference and attempt to synchronize with the master as the case with the algorithm TPSN or FTSP. In general the master node needs CPU resources in proportion of the number of slaves. Nodes with powerful CPUs are assigned to the master node.

- **Peer-to-Peer:** Most of the proposed protocols such as RBS and time diffusion protocol (TDP) are based on the structure of peer-to-peer any node can communicate directly with all other nodes in the network. This eliminates the risk of failure of a master node that would prevent synchronization of different configuration Peer-To-Peer offer more flexibility but also more difficult to control.

### 3.2. Probabilistic/Deterministic synchronization

- **Probabilistic synchronization:** In probabilistic time synchronization the offset value is measured probabilistically. The probabilistic approach is used because that a deterministic approach usually forces the synchronization protocol to perform more message transfers and simulate extra processing. It is very expensive in wireless environment where energy is scarce [27].

- **Deterministic synchronization:** Arvind (Arvind 1994) defines deterministic algorithms that guarantee an upper bound on the clock offset with certainty [27]. Most algorithms are deterministic in the literature like RBS [6] and TDP [26].

### 3.2. Clock Correction/untethered clocks

- **Clock Correction:** The clock function in memory is modified after each run of the time synchronization process, TPSN protocol uses this approach.

- **Untethered clocks:** Each clock live freely, but each node stores the data necessary to convert local time into the time base of each other.

### 3.3. Internal synchronization versus external synchronization

- **Internal synchronization:** The objective is to minimize the maximum difference between the readings of the clocks of the sensors. Can be performed in master slave or peer to peer.

- **External Synchronization**: A source of standard time such as Universal Time (UTC) is provided here. We do not need a global time base since we have an atomic clock that provides real-time in the real world usually called the reference time to be synchronized. Can not be done with Peer to Peer

### 3.4. Sender-to-receiver Versus receiver-to-receiver synchronisation

- **Sender-to-receiver:** Most existing methods are transmitter to a receiver by transmitting the values of the clocks, Therefore these methods are faced with variances of the delay message. The transmitting node periodically sends a message with its local time as a timestamp to the receiver and then the receiver synchronizes with the sender using the timestamp receiver from the sender.

    **1.** The transmitting node sends periodically a message with its local time as a timestamp by the receiver.

    2. The receiver then synchronizes with the sender using the timestamp received from the sender.

    3. The delay message between sender and receiver is calculated by measuring the total time to and from the moment the receiver requests Timestamp juice-only time he receives an actual response.

- **Receiver-to-Receiver:**

This approach exploits the property of the physical broadcast medium that if any two receivers receive the same message in a single-hop transmission (see below), they receive it at approximately the same time.

Table 1 classifies the various algorithms for clock synchronization, based on the analysis in this section.

**Table1: Classification based on synchronization issues**

| Algorithms | Synchronization Issues | | | |
| --- | --- | --- | --- | --- |
| | Master-Slave/Peer-to-Peer | Internal Vs External | Sender-to-receiver Vs Receiver-to-receiver | Clock correction |
| RBS 2002 | Peer to Peer | Both | Receiver-to-receiver | No |
| TPSN 2003 | Master | Both | Sender-to-receiver | Yes |
| FTSP 2004 | Master | Both | Sender-to-receiver | Yes |
| Miny-Sync 2003 | Peer to Peer | Internal | Sender-to-receiver | Yes |
| LTS 2003 | Master | Both | Sender-to-receiver | Yes |
| DMTS 2003 | Master | Both | Sender-to-receiver | No |
| TDP 2005 | Peer to Peer | Internal | Receiver-to-receiver | Yes |
| SLTP 2007 | Master | Both | Sender-to-receiver | Yes |
| GTSP 2009 | Master | Both | Sender-to-receiver | Yes |
| TSRT 2011 | Master | Both | Sender-to-receiver | No |

## 4. Existing Approach of Synchronization

In this following section we will examine the peculiarities of the most interesting algorithms in wireless sensor networks as shown in figure 1:

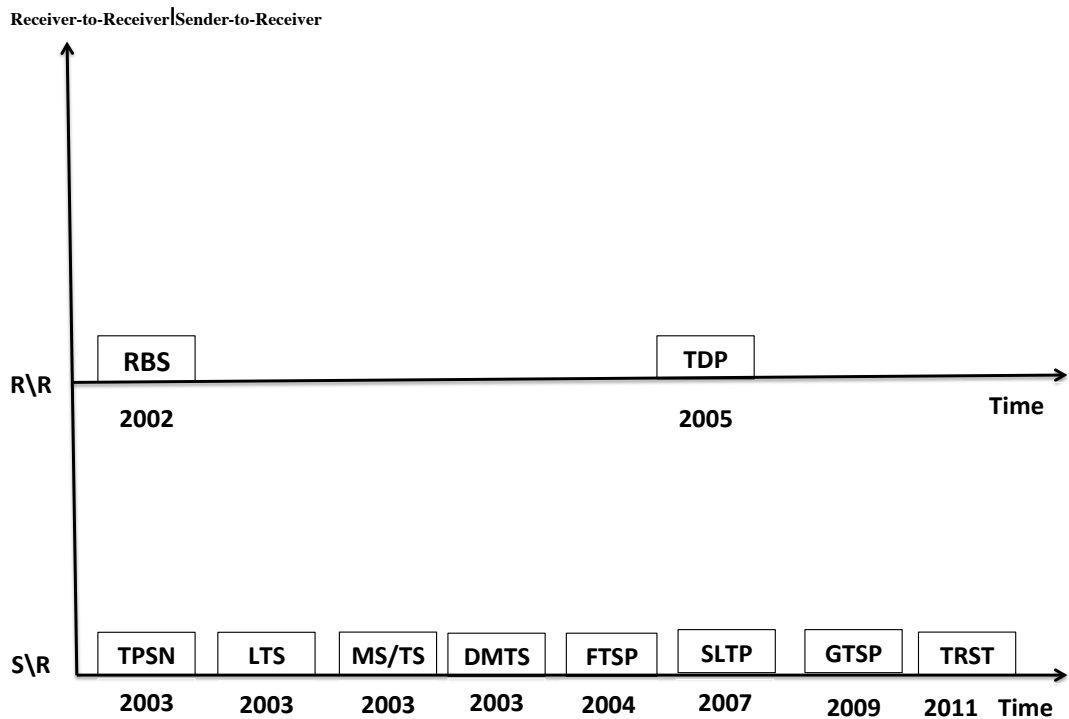

**Figure1:** Chronological Taxonomy of synchronization protocols

**4.1. Reference Broadcast Synchronization: RBS**

Elson and al. [6] have proposed the RBS approach which has been viewed as reference to several works in the same line of research (Synchronization Solution). The RBS synchronization mechanism is based on the exploitation of the nature of diffusion of wireless medium. With this property, the nodes in the transmission range of the same located in the intersection of two neighborhoods would be synchronized. Despite the advantages of elimination of major sources of indeterminism, transmitter receive the same message with a very low offset. Considering only the time for receipt of different receptors, the RBS protocol immediately eliminates two major sources of indeterminism involved in the transmission of messages, errors and the precision of synchronization that follows. The mechanism RBS has certain limitations: it requires that the reference receivers of messages transmitted by the transmitter to know the real time and the advanced channel time. The only sources of indeterminism that interfere in RBS synchronization is the propagation and receipt time which shall exchange times of receptions.

### 4.2. Timing-sync Protocol for Sensor Network: TPSN

Garnewel and al [5] have proposed an alternative approach to synchronization with the type Transmitter-receiver, TPSN is a hierarchical algorithm which works on two different phases:

The discovery and synchronization phase. In the first phase, we give a network node level. The node that initiates the synchronization is called the root node with the value of level zero neighbors with n hops (n=1, 2, 3, K) have the value of level n. This process continues until all neighbors attribute their levels. In second phase, a pair wise synchronization is performed along the edges of the hierarchical structure up to a total synchronization of the tree constructed with the message exchange mechanism.

### 4.3. Flooding Synchronisation Time Protocol: FTSP

The objective of FTSP [4] is to achieve a local synchronization with local participating nodes. Assuming that each node has local clock synchronization errors, and can communicate despite the lack of reliability the errors must be corrected with message exchange mechanism. FTSP synchronize time from a sender to multiple receivers which may be us- ing a single radio message. This mechanism could ensure high accuracy between two sensors and keep synchronized communication. Typically, WSN (Wireless Sensor Network) operates in larger areas than the radius of a node. Therefore, the FTSP synchronizes multi-hop nodes. The root node is the only selected and dynamic node that maintains the overall time for all other nodes to synchronize their clocks. The nodes form a Ad-hoc structure to transfer the total time from the root to all nodes that keep (save) the initial phase of the tree that is more robust against failures of links between nodes and the dynamic topology change.

### 4.4. Lightweight Tree-based Synchronization: LTS

Lightweight Tree-based synchronization [7] is different from other work [4], [5], [6] in the sense that its objective is not to maximize accuracy but to reduce the complexity of synchronization. Thus, the timing accuracy required is supposed to be given as a constraint and the main objective is to have a synchronization algorithm with minimal complexity to achieve given accuracy. The accuracy of the maximum time needed in sensor networks is relatively small (fractions of seconds) and we just have to use a synchronization scheme in the network. The two proposed algorithms for synchronizing multi-hop require nodes to be synchronized to some reference point(s) such as a sink node in the sensor network. Two ways are possible as follows.

#### 4.4.1. Centralized Multi-hop

It is a simple linear extension of single-hop synchronization. The basis of the algorithm is the construction of a spanning tree T to a Shallow including network nodes. In general whenever a new spanning tree is constructed each time the algorithm is executed. In order to synchronize the nodes in the tree, synchronizations are performed in pairs along the edges of T with synchronization centralized multi-hop. The reference node begins the synchronization mechanism with all the twigs mediation T. Then each child node is synchronized with the reference other children. This process continues until all nodes leaves of T reaches the algorithm ends when all leaf nodes are synchronized. The execution time of the algorithm is proportional to the depth of the tree.

#### 4.4.2. Distributed Multi-hop

It needs to build a spanning tree. This algorithm performs the synchronization of nodes in a distribution mode and doesn't use an overlay tree to direct synchronization pair. This algorithm is charged with the responsibility of synchronization from the reference node by the nodes themselves. The rate of synchronization of individual nodes can be determined using the same parameters as the reference nodes used in the central case.

### 4.5. Tiny-Sync and Mini-Sync

The two algorithms have common characteristics [8]: they are very tolerant of loss-message :

- They have limited storage and computation complexity
- They can be extended to all two-way communication of data.

#### 4.5.1. Tiny-Sync

It is based on observation. The data points obtained from the development of the estimation procedure are not all useful. Each data point consists of two constraints which are bounded by the offset and clock skew. At any time constraints only four are preserved instead of six constraints on the arrival of a new data point. The two new constraints are compared with existing data points and the four constraints resulting in the best estimates are selected by the calculation complexity of the algorithm.

### 4.5.2. Miny-Sync

Is an extension of Tiny-Sync it finds the optimal solution with increas- ing complexity. The idea is to prevent the algorithm being used by some data points has come to give strict limits. The authors develop a test to Scalable determine if there is a chance that a constraint could be useful. A constraint is eliminated if it is definitely unitils (points that do not change the solution).

### 4.6. Scalable Lightweight Time Synchronization Protocol for Wireless Sensor Network: SLTP

Uses the method of collecting (Clustering) and linear regression that reduces energy consumption of network [20]. SLTP works in two phases, phase one concerns the configuration for static and dynamic network in which determines the node group leader and members The second phase allows timing synchronization network after selecting the group leader and then the network initiates the synchronization.

### 4.7. Reference Based, Tree Structured Time Synchronization: TSRT

Proposed by Surendra Rahamatkar and Ajay Agarwal [12], The aim is to minimize the complexity of synchronization. TSRT has two main steps to synchronize the network, The building of an ad hoc tree structure is the first, the second used to synchronize the local clocks of sensors nodes followed by network evaluation phase. At the end of synchronization phase, the network realizes the wide synchronization of the local clocks of the participating nodes.

There are other interesting works by the problems of synchronization inherit all of the algorithms described above as:

- TSYNC [15]: the basic mechanism uses multi-channel nodes. It means that each node has two channels for control and clock. All nodes use the one channel for control; another channel for The latter manages the traffic of all network. The protocol has two version (centralized and decentralized). The first named RHS (Hierarchy Referencing Time Synchronization Protocol) is used to synchronize the whole network and the second is ITR (Individual-based Time Request Protocol) that allows each node to synchronize on demand. Each node that wants to be synchronized sends a request message to its parent RTI and this is repeated until the request message reaches the base station and the latter returns the clock through the channel clock to the node called.

- Li.Ming [16]: He defines a synchronization protocol based on Time Synchronization Protocol for WSNs spanning tree based on a hierarchical structure. In this protocol a first spanning tree of all nodes in the network is created. It is divided into multiple subtrees and each subtree is a set of child nodes. The sub tree are identified by the father-level node is level 0. The mechanism of synchronization subtree is currency in three phases.

- Delay Measurement Time Synchronization for Wireless Sensor Networks DMTS [18] bases on a master-slave synchronization, sender-receiver synchronization, and the approach of clock correction. The aims objetif of this protocol is to avoid the round-trip time. DMTS synchronize the transmitter and multiple receivers at the same time and requires fewer message transfers of RBS. Another advantage is their self-organization and dynamic behavior. The function of self-organization implies that the network topology may change from time to time. DMTS focuses on the scalability and flexibility, which means either to adapt or be insensitive to changes in network topology.

**5. Evaluation of Time Synchronization Algorithms**

In this section we will detail the different characteristics and synchronization problem of wireless sensor network during its operations. The usefulness or availability of sensor network is made to meet the needs of user requests by merging data from each sensor to provide a single result. To accomplish this task it becomes necessary for these sensors to agree on a concept like time. All active sensors (participants) can be wrapped in a common time scale either by synchronizing local clocks in each sensor either by transfer stamping. This section presents a comparison of various time synchronization algorithms for WSN's based on different principal factors such as the accuracy, the energy efficiency, the mobility and the complexity.

- **Energy efficiency:** This is an implicit requirement in most wireless networks in which this obligation must be executed and vary according to the demand. For example, in the case of sensor networks this requirement is strict forcing the nodes to sleep as often as possible and severely limiting the energy available for synchronization and other tasks. The main reason behind this constraint energy is the small size of the batteries sensors. This limits the amount of energy produced and stored.

- **Accuracy:** The accuracy is the measure of how time maintained within the network is confirmed to standard time. In other words, it is a measure of the precision of synchronization. A protocol with high precision ensures high accuracy in the case of absolute precision. This means that the time synchronized in the network does not deviate much from the external reference (eg, UTC). In the case of a relative accuracy means when a synchronized set of nodes is considered the maximum deviation of a clock node whose set is relatively small.

- **Computational complexity:** wireless networks often have limited physical capacity and energy constraints. So, the complexity of a synchronization protocol can take a practice protocol for many applications when we distinguish between the computational complexity of a protocol (for such execution and memory requirements) and message complexity (number of messages exchanged during synchronization).

- **Mobile networks:** In a mobile network, the sensors have the ability to move, and they connect with other sensors only when entering the geographical area of those sensors. The area of a mobile sen sor is the communication range up to which it can communicate and exchange messages successfully with other sensors. Romer [21] shows the need for a robust protocol, which can handle the frequent changes in network topology due to the mobility of the nodes. The change in topology is often a problem because it requires resynchronization of nodes and re computation of the neighborhoods or clusters.

**Table2: ALGORITHMS TAXONOMY PER SYNCHRONIZATION Issues [25]**

| | Application characterstics | | | |
|---|---|---|---|---|
| Algorithms | Energy efficiency | Mobility | Complexity | Accuracy |
| RBS 2002 | High | No | High | High |
| TPSN 2003 | High | No | Low | High |
| FTSP 2004 | High | Yes | High | High |
| Miny-Sync 2003 | High | No | Low | High |
| LTS 2003 | Low | Yes | Low | Average |
| DMTS 2003 | Very High | No | Low | High |
| TDP 2005 | Average | Yes | High | High |
| SLTP 2007 | High | Yes | Average | High |

| GTSP 2009 | High | No | Average | High |
| TSRT 2011 | Low | Yes | Average | High |

The Table-2 illustrates the pros and cons of the characteristics of 11 different algorithms [25], as well as the capability and reliability of each one. It is important to note that the recent works as well as SLTP have dealt with the problems of mobility and energy consumption unlike RBS for instance, which is mainly attributed to the evolve of the daily needs and exigent requirements of modern life. Also he table above summarizes, the efficiency of the most common solutions (algorithms) and Cause/Effect analyzes of the major glitches preventing the detectors proper functioning, and which among them managed to overcome many of these constraints. Many different network models have been proposed we started with the most robust solution against the big number of issues. FTSP proves good precision despite failures link and dynamic topology, it save the initial phase and establishing the tree with high energy efficiency with less resources then other solution in same categories like TPSN and RBS. Also FTSP prove a good results not only on a fixed networks hierarchy but updates it continuously, it supports network topology changes including mobile nodes. Another solution that provided high precision is the TPSN. Although it provided a good precision in order of 16.9μs, still the TPSN operates on a fixed infrastructure (hierarchical structure), structure increases as consumption increase. RBS or popular synchronization algorithm are characterized by ability to eliminate the uncertainty of sender by removing the sender from the critical path (way of sending) that give value of precision with a value of High energy consumption over RBS gives good results with the fixed architecture of networks. Another protocol proposed by which requires high power because the stations (nodes of Networks) used to discipline the local time of the nodes in the network, this value increases if working in a mobile network and characterized by a high complexity (i.e. the number of messages exchanged during synchronization). A third synchronization algorithm LTS is as important as the aforementioned solutions, but because of its high energy consumption it is not very effective as it requires a physical clock correction to perform on local clock of sensors while achieving synchronization. In addition, it is not recommended to use in mobile networks because it requires a hierarchical infrastructure with high mobility. The simulation results show that the accuracy of LTS is about 0.5 seconds. Like DMTS the Miny and Tiny sync is not applicable for the mobile sensor networks but characterized by low complexity and less power consumption as compared to other solutions. It is also noted that recent solutions like SLTP and TRST are studying the problem of energy consumption and mobility for the general development of all in the military field, for example sensors that can be mounted on the soldiers, and medical staff that requires some mobility. At the end of this section, it can be clearly seen that all solutions provided until now have in common is a good precision levels but requires high level of energy consumption, which opens the doors for more work and research on synchronization timing issues to come up with a solution that is very accurate, low in energy consumption, and cost effective. The following table summarizes the capability of each solution with the major criterion for the best timing as needed.

## 6. CONCLUSIONS

Energy efficiency is crucial because of the scale and application environments in which sensors are deployed. Wireless sensor networks offer great advantages for monitoring object movement and environmental properties but require some degree of synchronization to achieve the best results. However, providing clock synchronization among the sensor nodes, while designing and building such sensor networks remain a challenge.

This paper provides an in depth analysis of the most known existing clock synchronization algorithms based on a palette of factors like precision, accuracy, and complexity. The WSN algorithms presented here can help developers either in choosing an existing synchronization algorithm or in defining a new one that is best suited to the specific needs of a sensor-network application. Finally, the survey provides a valuable framework by which comparative evaluation of new and existing synchronization algorithms can be accomplished.

**Authors**

Salim El Khediri, a Ph.D student at the National School of Engineers of Sfax, Tunisia (Laboratory of Electrical and Information Technology) and in partnership collaboration education with CNAM laboratory CNAM (National Conservatory of Arts and Métiers Paris-France). Since 2009 working in the capacity of assistant teacher at the Faculty of Sciences Gafsa – Tunisia.

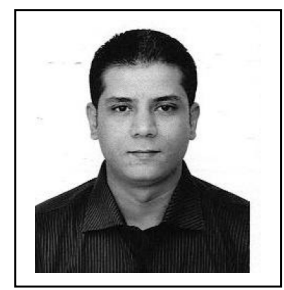